# Surface curvature and secondary vortices in steady dense shallow granular flows

**C. Gadal[1], C. G. Johnson[1], and J. M. N. T. Gray[1]**

[1]Department of Mathematics and Manchester Centre for Nonlinear Dynamics, University of Manchester, Oxford Road, Manchester M13 9PL, U

**Corresponding author:** C. Gadal, cyril.gadal@manchester.ac.uk

Dense granular flows exhibit both surface deformation and secondary flows due to the presence of normal stress differences. Yet, a complete mathematical modelling of these two features is still lacking. This paper focuses on a steady shallow dense flow down an inclined channel of arbitrary cross-section, for which asymptotic solutions are derived by using an expansion based on the flow shallowness combined with a second-order granular rheology. The leading order flow is uniaxial, with a streamwise velocity corresponding to a lateral juxtaposition of Bagnold profiles scaled by the varying flow depth. The correction at first order introduces two counter-rotating vortices in the plane perpendicular to the main flow direction (with downwelling in the centre), and an upward curve of the free surface. These solutions are compared to DEM simulations, which they match quantitatively. This result is then used together with laboratory experiments to infer measurements of the second-normal stress difference in dense dry granular flow.

## 1. Introduction

When set in motion, granular materials exhibit a range of intriguing behaviors that arise from complex interactions at the grain scale. One peculiar phenomenon is the upward curvature of the free surface of dense granular flows in certain configurations, such as open channels with frictional walls (Félix & Thomas 2004; Deboeuf *et al.* 2006; Takagi *et al.* 2011; McElwaine *et al.* 2012), non-rectangular channel geometries (see experiments in figure 1 and section 6) and Couette cells (Dsouza & Nott 2021; Cabrera & Polanía 2020). Grain-scale numerical simulations indicate that this surface deformation is accompanied by the existence of secondary counter-rotating vortices in the plane perpendicular to the main flow direction (Krishnaraj & Nott 2016; Dsouza & Nott 2021; Kim & Kamrin 2023).

These secondary flows have also been observed recently in experiments involving slow bulldozing of granular material using a conveyor belt (Einav *et al.* 2024). Interestingly, both surface curvature and vortices are enhanced in flows of elongated particles, supporting the idea that these effects stem from mechanisms occurring at the grain scale (Wortel *et al.* 2015; Fischer *et al.* 2016; Mohammadi *et al.* 2022).

In dense granular flows, the surface curvature and secondary flows are generally attributed to a non-zero, negative, second normal stress difference (between the vorticity and the gradient direction) resulting from an anisotropic distribution of the contacts at the grain scale (Nagy *et al.* 2020; Srivastava *et al.* 2020). This is more generally related to the broken codirectionality and coaxiality between the stress and strain rate tensors, as observed in grain-scale simulations (e.g. Campbell 1989; Silbert *et al.* 2001; Weinhart *et al.* 2013; Srivastava *et al.* 2020). Note however that both surface deformation and secondary vortices have also been reported in rapid chute flows. In that regime, they are usually attributed to fluctuations of the granular temperature, which also cause significant spatial variations in particle volume fraction (Forterre & Pouliquen 2001; Börzsönyi *et al.* 2009; Brodu *et al.* 2015).

Dense inertial granular flows are commonly described using the so-called $\mu(I)$ granular local rheology (GDR MiDi 2004; Jop *et al.* 2006). In this framework, the effective viscosity of the granular fluid depends on the local pressure due to the friction between the grains, but also on a dimensionless quantity, the inertial number $I$, which compares the time scale of macroscopic deformation with that of grain-scale rearrangements (GDR MiDi 2004). This constitutive law has been applied to a variety of flow configurations, including inclined channels, silo discharges and column collapses (e.g. Jop *et al.* 2006; Lagrée *et al.* 2011; Staron *et al.* 2014; Martin *et al.* 2017). However, the original $\mu(I)$ rheology presents several limitations. First, it is mathematically ill-posed at both low and high values of $I$ when combined with the canonical shape of the $\mu(I)$ curve (Barker *et al.* 2015). Partial regularization, using a specific shape of the $\mu(I)$ curve, is then be required to obtain grid-resolution convergent simulations below a critical inertial number (Barker *et al.* 2017, 2021; Maguire *et al.* 2024; Lloyd *et al.* 2025). Second, the original form of this rheology can not account for certain phenomena observed in experiments or grain-scale simulations. For instance, some flows can dilate and contract, motivating the development of compressible extensions in which the volume fraction $\phi$ also depends on the inertial number, in addition to the friction coefficient $\mu$ (GDR MiDi 2004; Pouliquen *et al.* 2006; Forterre & Pouliquen 2008). However, such compressible $\mu(I)$ models are generally ill-posed unless formulated under specific frameworks (Heyman *et al.* 2017; Barker *et al.* 2017; Schaeffer *et al.* 2019). Moreover, some flows exhibit non-local effects responsible for the slow creeping motions observed in heap flows far from the free surface (Komatsu *et al.* 2001) or coexisting adjacent static and flowing regions (e.g. Edwards & Gray 2015; Rocha *et al.* 2019), which lead to corresponding extensions of the $\mu(I)$ rheology (Pouliquen & Forterre 2009; Bouzid *et al.* 2015; Kamrin 2019).

To account for broken codirectionality and the existence of normal stress differences, the $\mu(I)$ rheology can also be extended to a second-order formulation, similar to the general form of a second-order fluid, also known as the Rivlin-Ericksen fluid of second grade (Rivlin & Ericksen 1997; McElwaine *et al.* 2012; Srivastava *et al.* 2020). Recently, this second-order rheology has also been extended to incorporate non-locality (Kim & Kamrin 2023). Using the local version of this second-order rheology, McElwaine *et al.* (2012) linked the upward surface curvature to the cross-stream gradient of the streamwise velocity, induced by friction on static levees present on the sides. However, the lack of a closure describing the flow behaviour at the interface between flowing and static regions prevented a complete resolution of the problem. Recently, Kim & Kamrin (2023) numerically solved the continuum mass and momentum equations using this second-order rheology and successfully reproduced the secondary flows observed in their grain-scale (DEM) simulations. However, rather than solving the complete free-boundary problem, they imposed spatial boundary conditions (including the free surface shape) directly extracted from the DEM simulations. Hence, a complete model of the entire flow structure in open channels remains elusive.

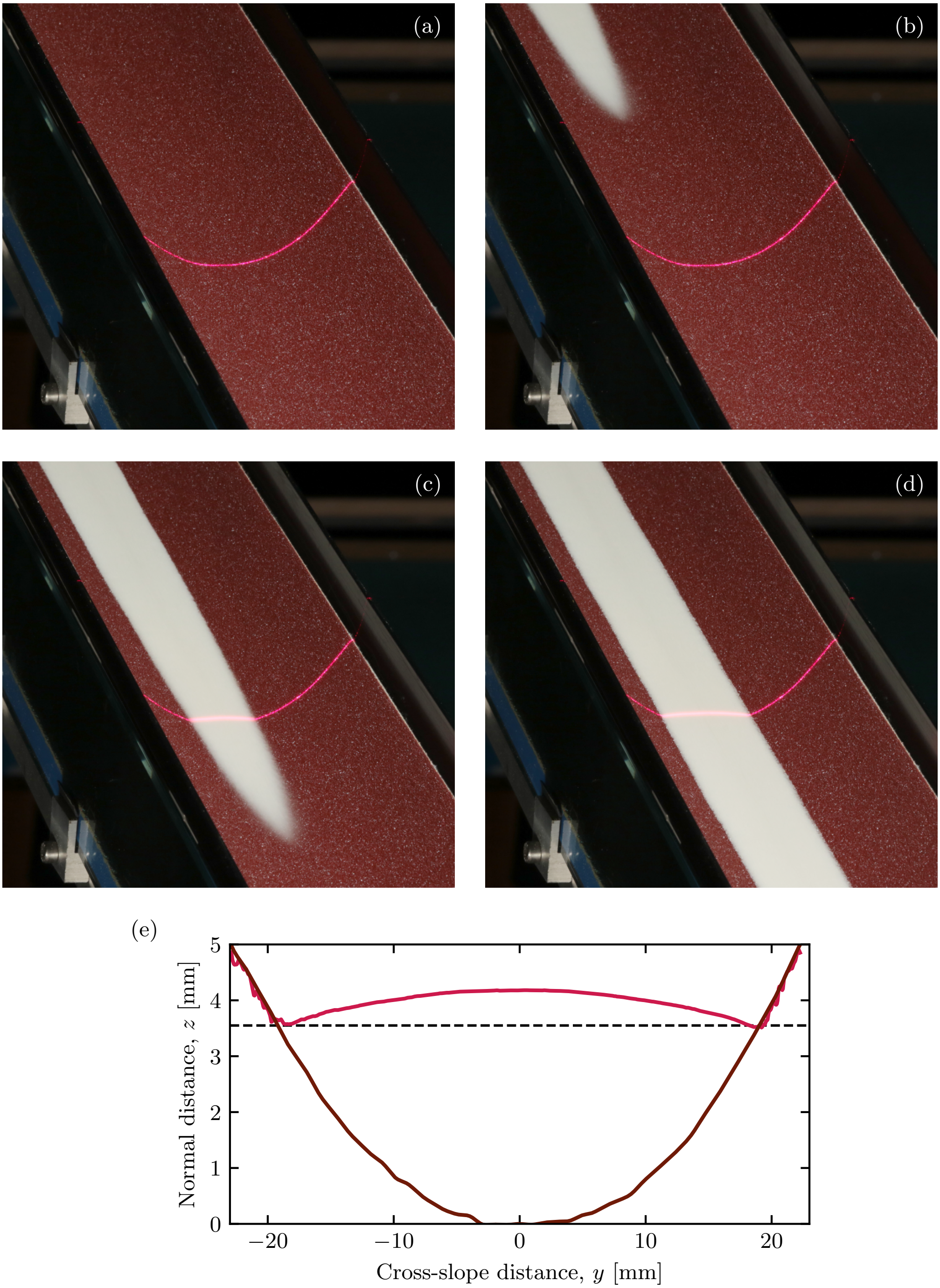

Figure 1. Dense flow of glass beads of diameter $d \in [125, 165]$ $\mu$m down an inclined channel with a parabolic cross-section in a laboratory experiment. (a–d) Snapshots of an experiment, featuring the empty channel, the propagating flow front and the steady uniform flow (see also movie 1 of the supplementary material). The channel is coated with a rough brown sandpaper, and the laser line shows the topography. (e) Topography laser measurements for an inclination of 26° and a flux of 51 g s$^{-1}$. The cross-section of the empty channel (corresponding to (a)) is shown by the brown line. The surface of the steady-state granular flow (corresponding to (d)) is shown by the red line, and is curved upward compared to the flat black dashed line. See section 6 for additional details on the experiments.

Surface deformation or secondary flows are also observed in other types of non-Newtonian fluids, such as viscoelastic fluids or dense suspensions (e.g. Maklad & Poole 2021). Dense suspensions, in particular, share several features with dense granular flows due to the frictional interactions between the particles forming the suspension (Guazzelli & Pouliquen 2018). In their case, the second normal stress difference has been measured using standard rheometer geometries (Singh & Nott 2003; Dbouk *et al.* 2013), but also using an alternative approach based on the measurement of the surface deformation in Weissenberg (rotating-rod) and tilted-trough geometries, in which confinement effects can be reduced and sensitivity improved (Boyer *et al.* 2011; Couturier *et al.* 2011; Dai *et al.* 2013). However, these approaches rely on a known relation between the normal stress difference and the surface deformation, which is currently lacking in the case of dense granular flows.

This paper starts by extending the approach of McElwaine *et al.* (2012) by considering a steady granular flow inside a tilted channel of general (but smooth) cross-section, and derives asymptotic solutions for the flow structure and surface deformation. These solutions are then compared to the results of grain scale (DEM) simulations, which they match quantitatively. Finally, laboratory experiments are performed, in which simple measurements of the surface deformation and the channel shape are used to infer measurements of the second normal stress differences using the results of the mathematical model.

## 2. Governing equations and second-order granular rheology

### 2.1. *System and governing equations*

The system considered here is sketched in figure 2. It consists fo a dense flow of particles, of diameter $d$ and intrinsic density $\rho_{\mathrm{p}}$, down an inclined channel of arbitrary cross-section. Let $Oxyz$ be a Cartesian coordinate system with the $x$-axis oriented along the slope, at an angle $\theta$ to the horizontal, and the $y$-axis and $z$-axis oriented in the cross-slope and upward normal directions, respectively. The flow has velocity components in the three directions, i.e. $\mathbf{u} = (u, v, w)$ and a bulk density $\rho = \phi\rho_{\mathrm{p}}$, where $\phi$ is the particle volume fraction.

The dynamics is then governed by the mass conservation equation

$$\frac{\partial \rho}{\partial t} + \boldsymbol{\nabla} \cdot (\rho \mathbf{u}) = 0, \tag{2.1}$$

and the momentum balance equation

$$\rho \frac{\partial \boldsymbol{u}}{\partial t} + \rho(\mathbf{u} \cdot \boldsymbol{\nabla})\mathbf{u} = \boldsymbol{\nabla} \cdot \boldsymbol{\sigma} + \rho \mathbf{g}, \tag{2.2}$$

where $\boldsymbol{\nabla}$ is the gradient operator, '$\cdot$' the dot product, $\boldsymbol{\sigma}$ the Cauchy stress tensor and $\mathbf{g} = g(\sin\theta, 0, -\cos\theta)$ the gravitational acceleration.

### 2.2. *The second-order granular rheology*

Following previous studies (McElwaine *et al.* 2012; Srivastava *et al.* 2020; Kim & Kamrin 2023), the granular flow is assumed to behave like a second-order fluid where the stress tensor can be constructed as a series of Rivlin-Ericksen tensors (Rivlin & Ericksen 1997).

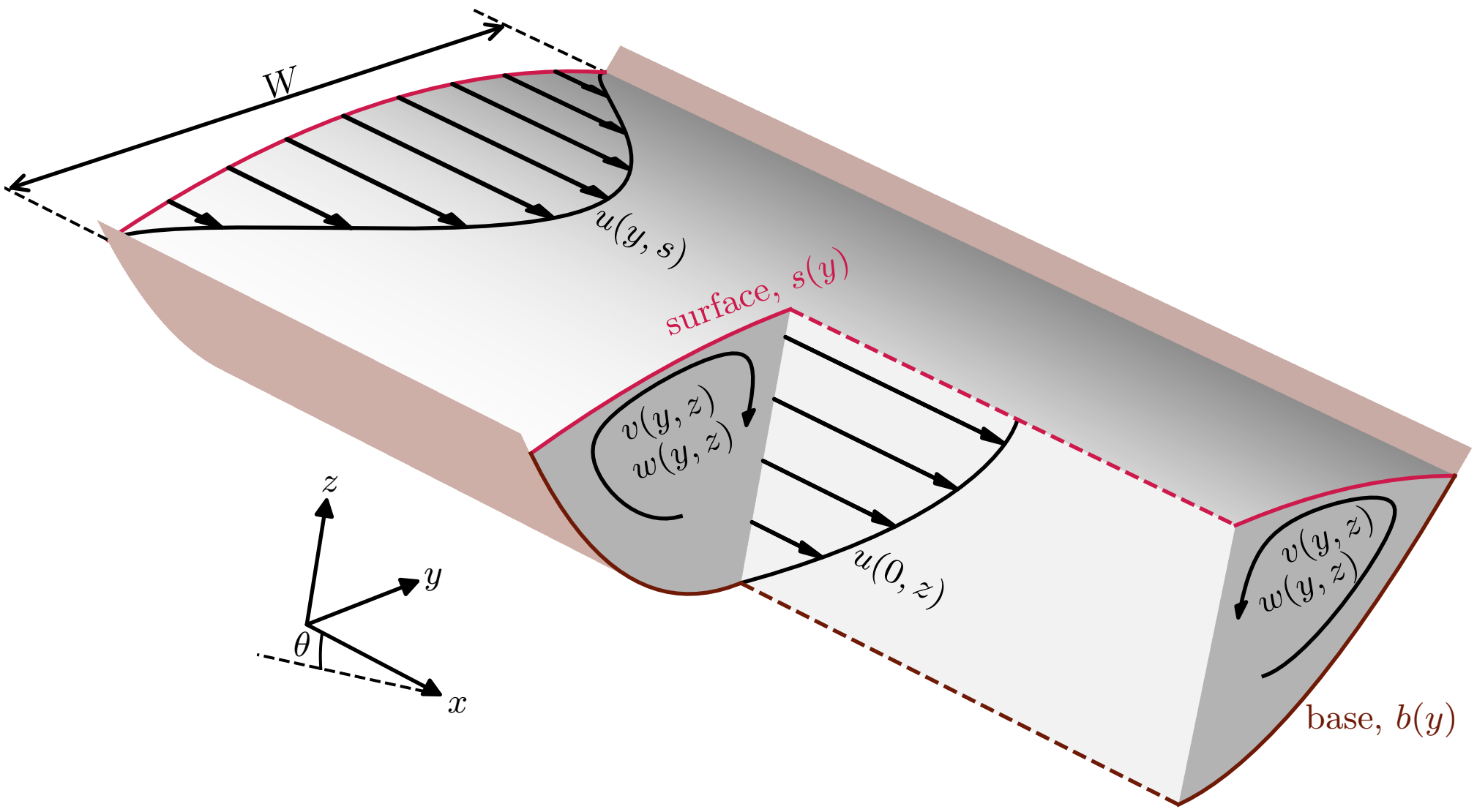

Figure 2. Conceptual sketch corresponding to the laboratory experiment shown in figure 1 and movie 1 of the supplementary material, showing the base shape, surface curvature and flow structure. The velocity $\mathbf{u} = (u, v, w)$ has a main component in the downslope direction $x$, and secondary flows $(v, w)$ in the cross-slope and normal directions. Note that the aspect ratio has been increased compared to the shallow flows considered here for the sake of clarity.

The strain rate and spin rate tensors, $\boldsymbol{D}$ and $\boldsymbol{W}$, are defined from the velocity gradient $\boldsymbol{\nabla}\mathbf{u}$ as

$$\boldsymbol{D} = \frac{1}{2}\left(\boldsymbol{\nabla}\mathbf{u} + \boldsymbol{\nabla}\mathbf{u}^T\right), \qquad \boldsymbol{W} = \frac{1}{2}\left(\boldsymbol{\nabla}\mathbf{u} - \boldsymbol{\nabla}\mathbf{u}^T\right). \tag{2.3}$$

The second-order Cauchy stress tensor is then

$$\begin{aligned}\boldsymbol{\sigma} = p\Bigg(-\boldsymbol{I} + \frac{\mu_1}{\|\boldsymbol{D}\|}\left[\boldsymbol{D} - \frac{1}{3}\mathrm{tr}\left(\boldsymbol{D}\right)\boldsymbol{I}\right] - \frac{\mu_2}{\|\boldsymbol{D}\|^2}\left[\boldsymbol{D}^2 - \frac{1}{3}\mathrm{tr}\left(\boldsymbol{D}^2\right)\boldsymbol{I}\right] \\ - \frac{\mu_3}{\|\boldsymbol{D}\|^2}\left[\frac{\partial \boldsymbol{D}}{\partial t} + (\mathbf{u}\cdot\boldsymbol{\nabla})\,\boldsymbol{D} + \boldsymbol{D}\boldsymbol{W} - \boldsymbol{W}\boldsymbol{D}\right]\Bigg),\end{aligned} \tag{2.4}$$

where $p \equiv -\mathrm{tr}\left(\boldsymbol{\sigma}\right)/3$ is the hydrostatic pressure, since all but the unit tensor $\boldsymbol{I}$ are deviatoric, and $\|\boldsymbol{D}\|$ is the second invariant of the strain-rate tensor:

$$\|\boldsymbol{D}\| = \sqrt{\frac{1}{2}\mathrm{tr}\left(\boldsymbol{D}^2\right)}. \tag{2.5}$$

Here, the definition of the shear rate is

$$\dot{\gamma} = 2\,\|\boldsymbol{D}\|\,. \tag{2.6}$$

The stress ratio coefficients $\mu_i$ may depend on any dimensionless parameters of the system. Neglecting the second-order terms, one recovers the so-called local '$\mu(I)$' rheology (Jop *et al.* 2006)

$$\boldsymbol{\sigma} = p\left[-\boldsymbol{I} + \mu_1\frac{\boldsymbol{D}}{\|\boldsymbol{D}\|}\right], \tag{2.7}$$

where the stress coefficient $\mu_1$ depends on a single dimensionless parameter, the inertial number

$$I = \frac{\dot{\gamma} d}{\sqrt{p/\rho_p}}, \tag{2.8}$$

corresponding to a ratio of timescales at the particle scale, $d/\sqrt{p/\rho_{\mathrm{p}}}$, and at the flow scale, $1/\dot{\gamma}$. Likewise, the second-order stress ratio coefficients have been shown to depend only on $I$ (Srivastava *et al.* 2020). Note that this is not the case for flows very close to rest (typically $I < 10^{-2}$), where these coefficients also depend on the granular temperature/fluidity as non-locality needs to be taken into account (Kim & Kamrin 2020, 2023). However, the inertial number is assumed to be large enough for these effects to be negligible.

Although various functional forms can be used to represent the $\mu_i(I)$ coefficients, the following derivations are made without prescribing any specific choice. However, in the first-order $\mu(I)$ rheology, the form of $\mu_1(I)$ determines the region of $I$ over which the rheology is well-posed (Barker *et al.* 2017). Correspondingly, the form of $\mu_1(I)$, $\mu_2(I)$ and $\mu_3(I)$ may influence the well-posedness of the second-order rheology (3.3). Analysis of well-posedness would be essential for studying time-dependent solutions of (3.3), but lies beyond the scope of this study, which focuses on steady-state solutions.

### 2.3. *Boundary conditions*

The kinematic and dynamic boundary conditions at the flow surface $s$ are

$$\frac{\partial s}{\partial t} + \mathbf{n}\cdot\mathbf{u} = 0, \quad z = s, \tag{2.9}$$

$$\boldsymbol{\sigma}\cdot\mathbf{n} = \mathbf{0}, \quad z = s, \tag{2.10}$$

where $\mathbf{n}$ is a unit vector normal to the flow surface. Additionally, there is no slip at the base $b$, such that

$$\mathbf{u} = \mathbf{0}, \quad z = b. \tag{2.11}$$

## 3. Assumptions and reduced problem

### 3.1. *Governing equations*

Motivated by the experiments shown in figure 1 and movie 1 of the supplementary material, the system is assumed to be steady in time and invariant along the $x$-axis, such that the flow velocity only depends on $y$ and $z$, and the flow height $h$ and channel cross-section $b$ only depend on $y$. The particle volume fraction $\phi$ and the bulk flow density ($\rho = \phi\rho_{\mathrm{p}}$) are also assumed to be constant and uniform (GDR MiDi 2004). Under those assumptions, the mass and momentum equations reduce to

$$\boldsymbol{\nabla}\cdot\mathbf{u} = 0, \tag{3.1}$$

$$\rho(\mathbf{u}\cdot\boldsymbol{\nabla})\mathbf{u} = \boldsymbol{\nabla}\cdot\boldsymbol{\sigma} + \rho\mathbf{g}. \tag{3.2}$$

The Cauchy stress tensor also simplifies to

$$\boldsymbol{\sigma} = p\left(-\boldsymbol{I} + \frac{\mu_1}{\|\boldsymbol{D}\|}\boldsymbol{D} - \frac{\mu_2}{\|\boldsymbol{D}\|^2}\left[\boldsymbol{D}^2 - \frac{1}{3}\mathrm{tr}\left(\boldsymbol{D}^2\right)\boldsymbol{I}\right] - \frac{\mu_3}{\|\boldsymbol{D}\|^2}\left[(\mathbf{u}\cdot\boldsymbol{\nabla})\,\boldsymbol{D} + \boldsymbol{D}\boldsymbol{W} - \boldsymbol{W}\boldsymbol{D}\right]\right), \tag{3.3}$$

as mass conservation implies that $\mathrm{tr}\,(\boldsymbol{D}) = 0$. In the case of simple planar shear flows, the second-order stress ratio coefficients can be expressed as functions of the scaled first and second normal stress differences, $N_1$ and $N_2$ (Srivastava *et al.* 2020; Kim & Kamrin 2023),

as

$$\mu_2 = -\left(N_2 + \frac{N_1}{2}\right), \tag{3.4}$$

$$\mu_3 = \frac{N_1}{4}. \tag{3.5}$$

Substituting the expressions of $\mu_2$ (3.4) and $\mu_3$ (3.5) into (3.3), the stress tensor then becomes

$$\begin{aligned}\boldsymbol{\sigma} = p\Bigg(&-\boldsymbol{I} + \frac{\mu_1}{\|\boldsymbol{D}\|}\boldsymbol{D} + \left[\frac{N_1}{2} + N_2\right]\frac{1}{\|\boldsymbol{D}\|^2}\left[\boldsymbol{D}^2 - \frac{1}{3}\mathrm{tr}\left(\boldsymbol{D}^2\right)\boldsymbol{I}\right] \\ &- \frac{N_1}{4\,\|\boldsymbol{D}\|^2}\left[(\mathbf{u}\cdot\nabla)\boldsymbol{D} + \boldsymbol{D}\boldsymbol{W} - \boldsymbol{W}\boldsymbol{D}\right]\Bigg),\end{aligned} \tag{3.6}$$

### 3.2. *Boundary conditions*

Under those assumptions, the boundary conditions at the flow surface (2.9) reduce to

$$v s_y = w, \quad z = s, \tag{3.7}$$

$$\sigma^{xy} s_y = \sigma^{xz}, \quad z = s, \tag{3.8}$$

$$\sigma^{yy} s_y = \sigma^{yz}, \quad z = s, \tag{3.9}$$

$$\sigma^{zy} s_y = \sigma^{zz}, \quad z = s. \tag{3.10}$$

The subscript notation is used for derivatives in this paper, such that $\partial X/\partial y = X_y$ for the $y$ coordinate, and similarly for the $z$ coordinate. Additionally, the flow depth is zero at the lateral edges

$$s = b, \quad y = \pm\frac{W}{2}, \tag{3.11}$$

where $W$ is the cross-stream horizontal extent of the flow. Finally, the conservation of the mass flux

$$Q = \rho\int_{-W/2}^{W/2}\int_{b(y)}^{s(y)} u(y,z)\mathrm{d}z\mathrm{d}y, \tag{3.12}$$

will be used to make comparisons between the mathematical model, the experiments, and the numerical simulations.

### 3.3. *Scalings and dimensionless equations*

Neglecting the normal stress differences, properties of downslope steady-state granular flows result from a balance between friction and gravity (e.g. GDR MiDi 2004), such that

$$\mu_1\mathcal{P} \sim \rho g\mathcal{H}\sin\theta, \tag{3.13}$$

where $\mathcal{H}$ and $\mathcal{P}$ are the characteristic height and pressure scales. As the pressure is hydrostatic

$$\mathcal{P} \sim \rho g\mathcal{H}\cos\theta, \tag{3.14}$$

this leads to $\mu_1(\mathcal{I}) \sim \tan\theta$, or equivalently to a characteristic inertial number

$$\mathcal{I} = \mu_1^{-1}(\tan\theta). \tag{3.15}$$

The inertial number is then controlled by the slope, but also the friction law of the considered granular media.

As the shear rate is dominated by the vertical velocity gradient $\mathcal{U}/\mathcal{H}$, (2.8) implies a Bagnold scaling for the velocity

$$\mathcal{U} \sim \mathcal{I}\frac{\mathcal{H}^{3/2}}{d}\sqrt{\phi g \cos\theta}. \tag{3.16}$$

The characteristic vertical and horizontal length scales, $\mathcal{H}$ and $\mathcal{W}$, can be determined by rewriting dimensionally the edge condition (3.11) and the flux conservation (3.12) as

$$\mathcal{H} = b\left(\frac{\mathcal{W}}{2}\right) \tag{3.17}$$

$$Q = \rho\mathcal{U}\mathcal{H}\mathcal{W} \tag{3.18}$$

The characteristic scales defined by (3.14), (3.15), (3.16), (3.17) and (3.18) suggests introducing dimensionless variables, indicated by the tilde, as

$$(z, h, y, W) = \mathcal{H}\left(\tilde{z}, \tilde{h}, \frac{\tilde{y}}{\epsilon}, \frac{\tilde{W}}{\epsilon}\right), \tag{3.19}$$

$$\mathbf{u} = \mathcal{U}\tilde{\mathbf{u}}, \tag{3.20}$$

$$(p, \boldsymbol{\sigma}) = \mathcal{P}(\tilde{p}, \tilde{\boldsymbol{\sigma}}), \tag{3.21}$$

$$I = \mathcal{I}\tilde{I}, \tag{3.22}$$

where $\epsilon = \mathcal{H}/\mathcal{W}$ accounts for the shallowness of the flow. Note that the inertial number $I$, although already dimensionless, is still scaled so that all leading-order quantities are of order $O(1)$. The dimensionless momentum and mass balance equations then become

$$\epsilon\tilde{v}_{\tilde{y}} + \tilde{w}_{\tilde{z}} = 0, \tag{3.23}$$

$$Fr^2\left(\epsilon\tilde{v}\tilde{u}_{\tilde{y}} + \tilde{w}\tilde{u}_{\tilde{z}}\right) = \tilde{\sigma}^{xy}_{\tilde{y}}\epsilon + \tilde{\sigma}^{xz}_{\tilde{z}} + \tan(\theta), \tag{3.24}$$

$$Fr^2\left(\epsilon\tilde{v}\tilde{v}_{\tilde{y}} + \tilde{w}\tilde{v}_{\tilde{z}}\right) = \tilde{\sigma}^{yy}_{\tilde{y}}\epsilon + \tilde{\sigma}^{yz}_{\tilde{z}}, \tag{3.25}$$

$$Fr^2\left(\epsilon\tilde{v}\tilde{w}_{\tilde{y}} + \tilde{w}\tilde{w}_{\tilde{z}}\right) = \tilde{\sigma}^{zy}_{\tilde{y}}\epsilon + \tilde{\sigma}^{zz}_{\tilde{z}} - 1, \tag{3.26}$$

where $Fr = \mathcal{U}/\sqrt{g\mathcal{H}\cos\theta} = \mathcal{I}\mathcal{W}\sqrt{\phi}\epsilon/d$ is the Froude number. The dimensionless velocity gradient is

$$\tilde{\nabla}\tilde{\mathbf{u}} = \begin{bmatrix} 0 & \epsilon\tilde{u}_{\tilde{y}} & \tilde{u}_{\tilde{z}} \\ 0 & \epsilon\tilde{v}_{\tilde{y}} & \tilde{v}_{\tilde{z}} \\ 0 & \epsilon\tilde{w}_{\tilde{y}} & \tilde{w}_{\tilde{z}} \end{bmatrix}, \tag{3.27}$$

from which all subsequent quantities (e.g. $\tilde{\boldsymbol{D}}$, $\tilde{\boldsymbol{\sigma}}$) can be computed. The dimensionless boundary conditions are

$$\epsilon\tilde{v}\tilde{s}_{\tilde{y}} = \tilde{w}, \quad \tilde{z} = \tilde{s}, \tag{3.28}$$

$$\epsilon\tilde{\sigma}^{xy}\tilde{s}_{\tilde{y}} = \tilde{\sigma}^{xz}, \quad \tilde{z} = \tilde{s}, \tag{3.29}$$

$$\epsilon\tilde{\sigma}^{yy}\tilde{s}_{\tilde{y}} = \tilde{\sigma}^{yz}, \quad \tilde{z} = \tilde{s}, \tag{3.30}$$

$$\epsilon\tilde{\sigma}^{zy}\tilde{s}_{\tilde{y}} = \tilde{\sigma}^{zz}, \quad \tilde{z} = \tilde{s}, \tag{3.31}$$

and the conservation of the total flux (3.12) becomes

$$\int_{-\tilde{W}/2}^{\tilde{W}/2}\int_{\tilde{b}(\tilde{y})}^{\tilde{s}(\tilde{y})} \tilde{u}(\tilde{y}, \tilde{z})\mathrm{d}\tilde{z}\mathrm{d}\tilde{y} = 1. \tag{3.32}$$

Finally, the inertial number is

$$\tilde{I} = \frac{\tilde{\dot{\gamma}}}{\sqrt{\tilde{p}}}. \tag{3.33}$$

## 4. Asymptotic solution for a shallow flow

Here, the shallowness of the flow is exploited to perform an asymptotic expansion of the system of equations. Hence, the flow is assumed to have a typical depth $\mathcal{H}$ much smaller than its width $\mathcal{W}$, such that $\epsilon = \mathcal{H}/\mathcal{W} \ll 1$. The flow is then assumed to be described by a streamwise base flow plus a small correction as

$$\tilde{\mathbf{u}} = \tilde{u}^{(0)}\mathbf{e}_x + \epsilon \begin{bmatrix} \tilde{u}^{(1)} \\ \tilde{v}^{(1)} \\ \tilde{w}^{(1)} \end{bmatrix} + \epsilon^2 \begin{bmatrix} \tilde{u}^{(2)} \\ \tilde{v}^{(2)} \\ \tilde{w}^{(2)} \end{bmatrix} + \dots, \tag{4.1}$$

where the superscripts $^{(i)}$ denote quantities of order $i$, respectively. The leading order of the mass balance equation (3.23) implies directly that $\tilde{w}^{(1)} = 0$. Similarly, other quantities can be expanded, including the rheological coefficients, as

$$(\tilde{\boldsymbol{\sigma}}, \tilde{p}, \tilde{I}) = (\tilde{\boldsymbol{\sigma}}^{(0)}, \tilde{p}^{(0)}, \tilde{I}^{(0)}) + \epsilon(\tilde{\boldsymbol{\sigma}}^{(1)}, \tilde{p}^{(1)}, \tilde{I}^{(1)}), \tag{4.2}$$

$$(\tilde{s}, \tilde{W}) = (\tilde{s}^{(0)}, \tilde{W}^{(0)}) + \epsilon(\tilde{s}^{(1)}, \tilde{W}^{(1)}), \tag{4.3}$$

$$(\mu_1, N_1, N_2) = (\mu_1^{(0)}, N_1^{(0)}, N_2^{(0)}) + \epsilon(\mu_1^{(1)}, N_1^{(1)}, N_2^{(1)}). \tag{4.4}$$

These are only expanded up to the linear order, as higher orders will not be involved in the expressions derived later.

### 4.1. *Base state velocity field*

Solutions for the base state are derived by keeping only the terms of order zero. The shear stress tensor reduces to

$$\tilde{\boldsymbol{\sigma}}^{(0)} = \begin{bmatrix} -A\tilde{p}^{(0)} & 0 & \mu_1^{(0)}\tilde{p}^{(0)} \\ 0 & -B\tilde{p}^{(0)} & 0 \\ \mu_1^{(0)}\tilde{p}^{(0)} & 0 & -C\tilde{p}^{(0)} \end{bmatrix}, \tag{4.5}$$

where

$$A = 1 - \frac{1}{3}\left(2N_1^{(0)} + N_2^{(0)}\right), \tag{4.6}$$

$$B = 1 + \frac{1}{3}\left(N_1^{(0)} + 2N_2^{(0)}\right), \tag{4.7}$$

$$C = 1 + \frac{1}{3}\left(N_1^{(0)} - N_2^{(0)}\right), \tag{4.8}$$

The first (between flow and gradient direction, $x$ and $z$) and second (between gradient and vorticity direction, $z$ and $y$) normal stress differences are then

$$\tilde{\sigma}^{xx(0)} - \tilde{\sigma}^{zz(0)} = N_1^{(0)}\tilde{p}^{(0)}, \tag{4.9}$$

$$\tilde{\sigma}^{zz(0)} - \tilde{\sigma}^{yy(0)} = N_2^{(0)}\tilde{p}^{(0)}, \tag{4.10}$$

as expected from the change of variables introduced at the end of section 3.1.

The mass balance equation and the $y$ momentum equation are trivially satisfied. The remaining $x$ and $z$ momentum equations then reduce to

$$\mu_1^{(0)} \tilde{p}_{\tilde{z}}^{(0)} = -\tan\theta, \tag{4.11}$$

$$-C\tilde{p}_{\tilde{z}}^{(0)} = 1. \tag{4.12}$$

Taking the ratio of (4.11) and (4.12) leads to an implicit equation for $\tilde{I}^{(0)}$ (recall that both $\mu_1^{(0)}$ and $C$, through $N_1$ and $N_2$, are functions of $\tilde{I}^{(0)}$)

$$\frac{\mu_1^{(0)}}{C} = \tan\theta. \tag{4.13}$$

This shows that, at leading order, the inertial number is constant across the whole flow, which is typical for dense granular flows on inclined planes (e.g. GDR MiDi 2004). It can then be computed analytically or numerically depending on the expressions of $\mu_1$, $N_1$ and $N_2$.

The shear rate, which simplifies here to $\tilde{u}_{\tilde{z}}^{(0)}$, is then selected from its relationship with the pressure (3.33) as

$$\tilde{u}_{\tilde{z}}^{(0)} = \tilde{I}^{(0)} \sqrt{\tilde{p}^{(0)}}. \tag{4.14}$$

The pressure can be obtained from (4.12)

$$\tilde{p}^{(0)} = \frac{1}{C}(\tilde{s}^{(0)} - \tilde{z}), \tag{4.15}$$

taking into account the dynamic boundary condition in the $z$-direction (3.31), which reduces to a vanishing pressure at the surface, $\tilde{p}(\tilde{s}^{(0)}) = 0$. By integrating vertically (4.14) and taking into account the no-slip condition at the bottom (2.11), a Bagnold velocity profile is recovered

$$\tilde{u}^{(0)} = \frac{2}{3\sqrt{C}} \tilde{I}^{(0)} \left[ \left(\tilde{s}^{(0)} - \tilde{b}\right)^{\frac{3}{2}} - \left(\tilde{s}^{(0)} - \tilde{z}\right)^{\frac{3}{2}} \right]. \tag{4.16}$$

At the leading order, the flow corresponds to a juxtaposition of non-interacting Bagnold velocity profiles, scaled by the cross-stream varying flow depth; that is, cross-slope stresses do not play a role (see figure 3). The normal stress differences contribute to decelerating the flow by modifying the vertical pressure distribution, since $C$ is slightly larger than unity according to the typical values of $N_1$ and $N_2$ found in DEM simulations (Srivastava *et al.* 2020; Kim & Kamrin 2023). In practice, this effect is unlikely to be measurable, as either both $\|N_1\|/3 \ll 1$ and $\|N_2\|/3 \ll 1$, or otherwise $N_1 \approx N_2$, such that $C \approx 1$ in any case (see equation 4.8).

To fully describe the base state, an equation describing the flow surface $s^{(0)}$ is required, which will come from looking at the equations at the next order.

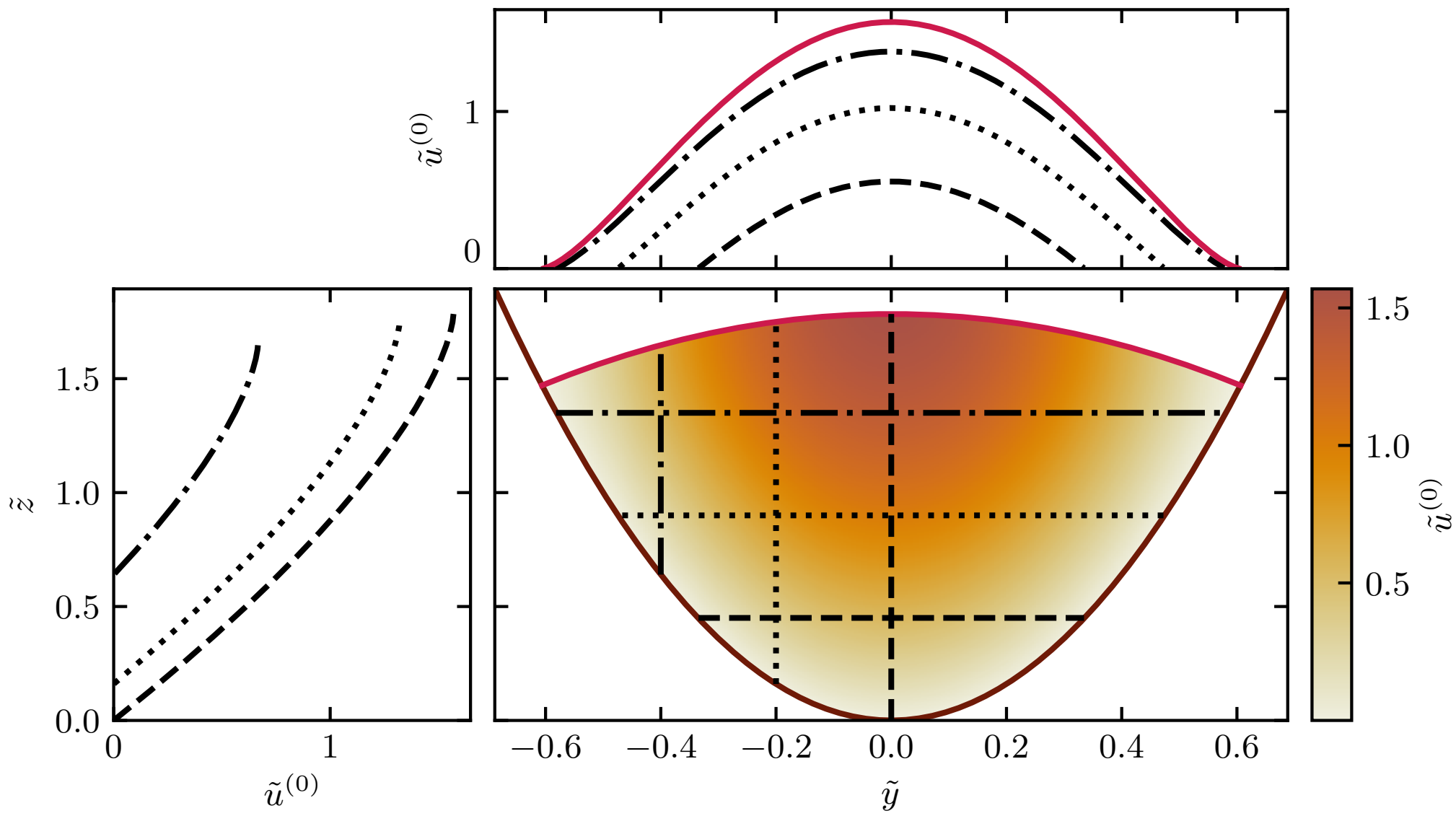

Figure 3. Base streamwise velocity $\tilde{u}^{(0)}$ computed from (4.16) in the case of a parabolic channel. The centre panel shows the velocity in colorscale in the plane $(\tilde{y}, \tilde{z})$. The red and brown lines represent the flow surface and channel base, respectively. The top and left panels show horizontal and normal velocity profiles, as annotated by the solid and dashed lines on the centre panel. In the top panel, the red curve shows the streamwise velocity along the surface, $\tilde{u}^{(0)}(\tilde{y}, \tilde{s}^{(0)})$.

### 4.2. *Linear correction and secondary flows*

As the system of equations is now balanced at the leading order, only terms of order $O(\epsilon)$ are kept. The relevant first-order corrections to the shear stress tensor are

$$\tilde{\sigma}^{xz(1)} = \mu_1^{(1)} \tilde{p}^{(0)} + \mu_1^{(0)} \tilde{p}^{(1)}, \tag{4.17}$$

$$\tilde{\sigma}^{yz(1)} = \frac{\tilde{p}^{(0)}}{\tilde{u}_{\tilde{z}}^{(0)}} \left( N_2^{(0)} \tilde{u}_{\tilde{y}}^{(0)} + \mu_1^{(0)} \tilde{v}_{\tilde{z}}^{(1)} \right), \tag{4.18}$$

$$\tilde{\sigma}^{zz(1)} = \frac{N_2 - N_1}{3} \tilde{p}^{(0)} - C\tilde{p}^{(1)}, \tag{4.19}$$

which lead to the momentum equations

$$\tilde{\sigma}_{\tilde{y}}^{xy(0)} + \tilde{\sigma}_{\tilde{z}}^{xz(1)} = \left[ \mu_1^{(1)} \tilde{p}^{(0)} + \mu_1^{(0)} \tilde{p}^{(1)} \right]_{\tilde{z}} = 0, \tag{4.20}$$

$$\tilde{\sigma}_{\tilde{y}}^{yy(0)} + \tilde{\sigma}_{\tilde{z}}^{yz(1)} = -B\tilde{p}_{\tilde{y}}^{(0)} + \left[ \frac{\tilde{p}^{(0)}}{\tilde{u}_{\tilde{z}}^{(0)}} \left( \mu_1^{(0)} \tilde{v}_{\tilde{z}}^{(1)} + N_2^{(0)} \tilde{u}_{\tilde{y}}^{(0)} \right) \right]_{\tilde{z}} = 0, \tag{4.21}$$

$$\tilde{\sigma}_{\tilde{y}}^{zy(0)} + \tilde{\sigma}_{\tilde{z}}^{zz(1)} = \left[ C\tilde{p}^{(1)} + \frac{N_1 - N_2}{3} \tilde{p}^{(0)} \right]_{\tilde{z}} = 0, \tag{4.22}$$

and the mass balance equation

$$\tilde{v}_{\tilde{y}}^{(1)} + \tilde{w}_{\tilde{z}}^{(2)} = 0 \tag{4.23}$$

Noting that $\tilde{p}_{\tilde{y}}^{(0)} = \tilde{s}_{\tilde{y}}^{(0)}/C$ is independent of $\tilde{z}$ and that the pressure vanishes at the surface, the $y$-momentum equation (4.21) can be integrated vertically from the surface to

obtain

$$\tilde{v}_{\tilde{z}}^{(1)} = \frac{1}{\mu_1^{(0)}} \left[ \frac{B}{C} \tilde{s}_{\tilde{y}}^{(0)} (\tilde{z} - \tilde{s}^{(0)}) \frac{\tilde{u}_{\tilde{z}}^{(0)}}{\tilde{p}^{(0)}} - N_2^{(0)} \tilde{u}_{\tilde{y}}^{(0)} \right], \tag{4.24}$$

This equation can be integrated vertically from the base, where the velocity vanishes, leading to

$$\tilde{v}^{(1)} = \frac{\tilde{I}^{(0)}}{3\mu_1^{(0)}\sqrt{C}} \left( 3N_2^{(0)} \sqrt{\tilde{s}^{(0)} - \tilde{b}}\, (\tilde{z} - \tilde{b}) \left(\tilde{b}_{\tilde{y}} - \tilde{s}_{\tilde{y}}^{(0)}\right) - 2C\tilde{s}_{\tilde{y}}^{(0)} \left[ \left(\tilde{s}^{(0)} - \tilde{b}\right)^{\frac{3}{2}} - \left(\tilde{s}^{(0)} - \tilde{z}\right)^{\frac{3}{2}} \right] \right). \tag{4.25}$$

Combining this expression for the cross-stream velocity and the mass balance equation (4.23) gives an expression for the normal velocity

$$\tilde{w}^{(2)} = \frac{\tilde{I}^{(0)}}{\mu_1^{(0)} \sqrt{C(\tilde{s}^{(0)} - \tilde{b})}} \left( \frac{N_2}{4} \mathcal{A} + \frac{C}{15} \mathcal{B} \right), \tag{4.26}$$

where

$$\begin{aligned} \mathcal{A} = \tilde{b} \Big[ \tilde{b} \left(\tilde{b}_{\tilde{y}} - \tilde{s}_{\tilde{y}}^{(0)}\right)^2 + 2\left(\tilde{b} - \tilde{s}^{(0)}\right) \left(\tilde{b} \left[\tilde{b}_{\tilde{y}\tilde{y}} - \tilde{s}_{\tilde{y}\tilde{y}}^{(0)}\right] + 2\tilde{b}_{\tilde{y}} \left[\tilde{b}_{\tilde{y}} - \tilde{s}_{\tilde{y}}^{(0)}\right]\right) \Big] \\ + \tilde{z} \Big[ (\tilde{z} - 2\tilde{b}) \left(\tilde{b}_{\tilde{y}} - \tilde{s}_{\tilde{y}}^{(0)}\right)^2 + 2\left(\tilde{b} - \tilde{s}^{(0)}\right) \left(2\tilde{b}_{\tilde{y}} \left[\tilde{s}_{\tilde{y}}^{(0)} - \tilde{b}_{\tilde{y}}\right] + [\tilde{z} - 2\tilde{b}] \left[\tilde{b}_{\tilde{y}\tilde{y}} - \tilde{s}_{\tilde{y}\tilde{y}}^{(0)}\right]\right) \Big], \end{aligned} \tag{4.27}$$

and

$$\begin{aligned} \mathcal{B} = \sqrt{\tilde{s}^{(0)} - \tilde{b}} \Big( 4\tilde{s}_{\tilde{y}\tilde{y}}^{(0)} \left[\tilde{s}^{(0)} - \tilde{z}\right]^{5/2} + 10 \left(\tilde{s}_{\tilde{y}}^{(0)}\right)^2 \left(\tilde{s}^{(0)} - \tilde{z}\right)^{3/2} \\ -5\tilde{z}\sqrt{\tilde{s}^{(0)} - \tilde{b}} \left[ 3\tilde{b}_{\tilde{y}} \tilde{s}_{\tilde{y}}^{(0)} + 2\tilde{s}_{\tilde{y}\tilde{y}}^{(0)} \left(\tilde{b} - \tilde{s}^{(0)}\right) - 3\left(\tilde{s}_{\tilde{y}}^{(0)}\right)^2 \right] \Big) \\ - \left(\tilde{s}^{(0)} - \tilde{b}\right) \left( 2\left[\tilde{s}^{(0)} - \tilde{b}\right] \left[ \tilde{s}_{\tilde{y}\tilde{y}}^{(0)} \left(2\tilde{s}^{(0)} + 3\tilde{b}\right) + 5\left(\tilde{s}_{\tilde{y}}^{(0)}\right)^2 \right] + 15\tilde{b}\tilde{s}_{\tilde{y}}^{(0)} \left[\tilde{s}_{\tilde{y}}^{(0)} - \tilde{b}_{\tilde{y}}\right] \right). \end{aligned} \tag{4.28}$$

The first non-zero corrections to the horizontal and normal velocities, $\tilde{v}^{(1)}$ and $\tilde{w}^{(2)}$, only depend on the quantities at the leading order. Computing first-order corrections to quantities that are already non-zero at the leading order, such as the pressure or the streamwise velocity, is then left for future work.

### 4.3. *Flow surface deformation*

At the linear order, the kinematic boundary condition (3.28) becomes

$$\tilde{v}^{(1)} \tilde{s}_{\tilde{y}}^{(0)} = \tilde{w}^{(2)}, \quad \tilde{z} = \tilde{s}^{(0)}. \tag{4.29}$$

Integrating the mass balance equation (4.23) and using Leibniz's integral rule leads to

$$\frac{\partial}{\partial \tilde{y}} \left( \int_{\tilde{b}}^{\tilde{s}^{(0)}} \tilde{v}^{(1)} \mathrm{d}\tilde{y} \right) - \left[ \tilde{v}^{(1)} \tilde{s}_{\tilde{y}}^{(0)} - \tilde{w}^{(2)} \right]_{\tilde{b}}^{\tilde{s}^{(0)}} = 0. \tag{4.30}$$

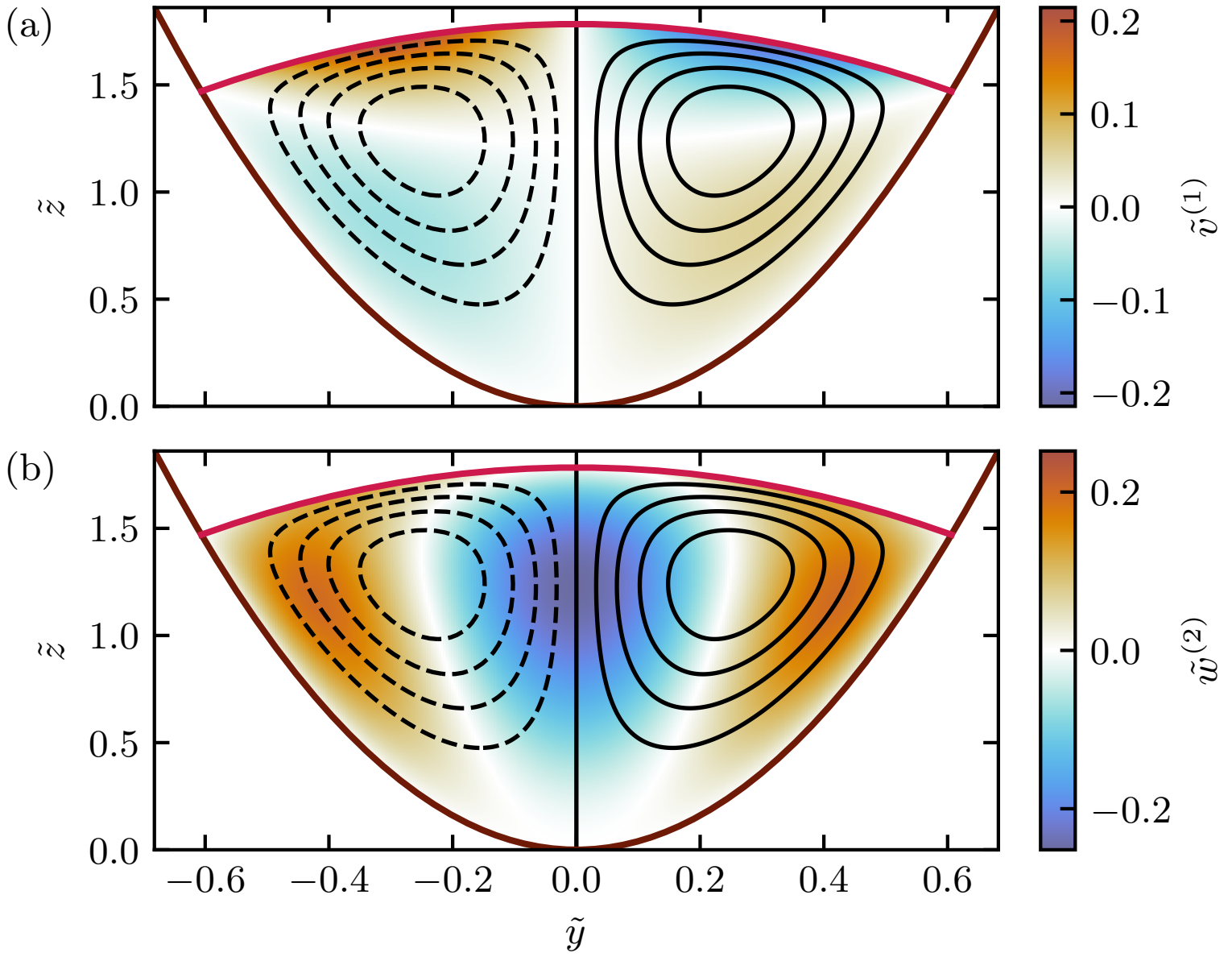

Figure 4. Secondary flows induced at the first order in the case of a parabolic channel. (a) Cross-stream velocity $\tilde{v}^{(1)}$ using (4.25) (b) Normal velocity $\tilde{w}^{(2)}$ using (4.26). In both subplots, the red and brown lines represent the flow surface and channel base, respectively. The black lines are streamlines, oriented clockwise (plain lines) and anti-clockwise (dashed lines).

Using the kinematic boundary condition (4.29) and no-slip boundary condition (2.11) simplifies the previous equation into

$$\frac{\partial}{\partial \tilde{y}}\left(\int_{\tilde{b}}^{\tilde{s}^{(0)}} \tilde{v}^{(1)} \mathrm{d}\tilde{y}\right) = 0, \tag{4.31}$$

implying that the cross-stream flux is constant in the cross-stream direction. As there is no flux through the lateral edges, at $\tilde{y} = \pm\tilde{W}/2$, the cross-stream flux is zero everywhere

$$\int_{\tilde{b}}^{\tilde{s}^{(0)}} \tilde{v}^{(1)} \mathrm{d}\tilde{y} = 0. \tag{4.32}$$

Combining this with the expression of $\tilde{v}^{(1)}$ obtained previously, (4.25), leads to a linear relation between the surface and the bottom gradients, which can be integrated as

$$\tilde{s}^{(0)} = \frac{5N_2^{(0)}}{5N_2^{(0)} + 4C}\left[\tilde{b} - \tilde{b}(0)\right] + \tilde{s}^{(0)}(0). \tag{4.33}$$

Finally, the problem can be closed by determining $\tilde{s}^{(0)}(0)$ and $\tilde{W}^{(0)}$ from the mass flux conservation and the boundary condition at the lateral edges

$$\int_{-\frac{\tilde{W}^{(0)}}{2}}^{\frac{\tilde{W}^{(0)}}{2}} \int_{\tilde{b}}^{\tilde{s}^{(0)}} \tilde{u}^{(0)} \mathrm{d}\tilde{z}\mathrm{d}\tilde{y} = 1 \tag{4.34}$$

$$\tilde{s}^{(0)} = \tilde{b}, \quad \tilde{y} = \pm\frac{\tilde{W}}{2}, \tag{4.35}$$

Hence, (4.33) shows that the free surface deformation is a leading order effect in the flow aspect ratio $\epsilon$, whose amplitude is directly related to the magnitude of the scaled second normal stress difference. The sign of the second normal stress difference determines if the surface deforms upward or downward. For granular material, $N_2$ is negative (compressive stresses in the $y$-direction are larger than in the $z$-direction), leading to an upward deformation of the surface (see figures 3 and 4) in agreement with the experimental observations (see figure 1 and section 6), numerical simulations (see section 5) and previous studies (McElwaine *et al.* 2012; Kim & Kamrin 2023). Importantly, this also determines the direction of rotation of the vortices, downwelling in the centre and going up on the sides (figure 4). This does not contradict the numerical results of Kim & Kamrin (2023), where the flow is upwelling in the centre, as their geometry is different and, in particular, far from shallow. This could change the balance between the cross-stream pressure gradient and the term induced by the second normal stress difference acting in the $y$-momentum equation (4.21). Finally, the absolute value of $N_2$ determines the magnitude of the surface deformation and the strength of the vortices. According to DEM simulations available in the literature (Srivastava *et al.* 2020; Kim & Kamrin 2023), these effects are therefore expected to become stronger as the inertial number of the flow increases, for example by increasing the channel inclination $\theta$. This is observed in laboratory experiments, as shown and discussed in section 6.

## 5. Model validation against DEM simulations

Due to the opacity of granular materials, measuring internal velocity fields remains a challenge in laboratory experiments. In order to verify the validity of the solution derived in section 4, numerical simulations are carried out at the grain scale using a Discrete Element Method (DEM) from open-source molecular dynamics code LAMMPS (Thompson *et al.* 2022).

### 5.1. *Numerical set-up*

The simulations consist in the dense flow of three-dimensional frictional spheres down an inclined rough parabolic channel. Details of the model, including contact forces and chosen parameters, are given in Appendix A. To achieve a proper comparison with the continuum modelling, the flow height in the centre is kept to around 25d to minimize finite size effects while keeping computation time reasonable. Additionally, the flow is kept shallow by having $\mathcal{H}/\mathcal{W} \simeq 0.09$. A typical DEM simulation is shown in movie 2 of the supplementary material.

The DEM simulations provide instantaneous information about the grain velocities and mechanical stresses, which are turned into averaged continuum fields following the method described in Appendix A. In particular, this gives access to the velocity and pressure, but also to the spatial distribution of the inertial number and friction coefficients (and as such normal stress differences), allowing direct comparison with the predictions of the model derived in sections 2 and 4.

### 5.2. *Results*

The continuum model in sections 2 and 4 is derived under the assumption of a constant volume fraction across the whole domain. As shown by figure 5a, this is verified in the DEM simulations, where the $\phi$ fluctuates in the domain around a constant value $\phi \in [0.53, 0.55]$ (depending on the slope angle), supporting the use of an incompressible rheology for dense granular flows. These fluctuations mostly come from layering at the sub-particle scale of particles close to the base, an effect present despite the polydispersity and the randomness of the roughness of the base as observed in previous studies (Weinhart *et al.* 2013). In addition, it also predicts the inertial number (and hence the friction coefficients and normal stress differences) to be constant at the leading order (see 4.13). As shown by figure 5b–e, this is also mainly the case in the DEM simulations, although some discrepancies exist. The inertial number is slightly larger at the edges and close to the surface, where the volume fraction is smaller and the shallow flow approximation breaks down. The first (resp. second) normal stress difference field exhibits a larger spatial structure, being smaller (resp. larger) at the top and at the edges than in the centre and at the bottom.

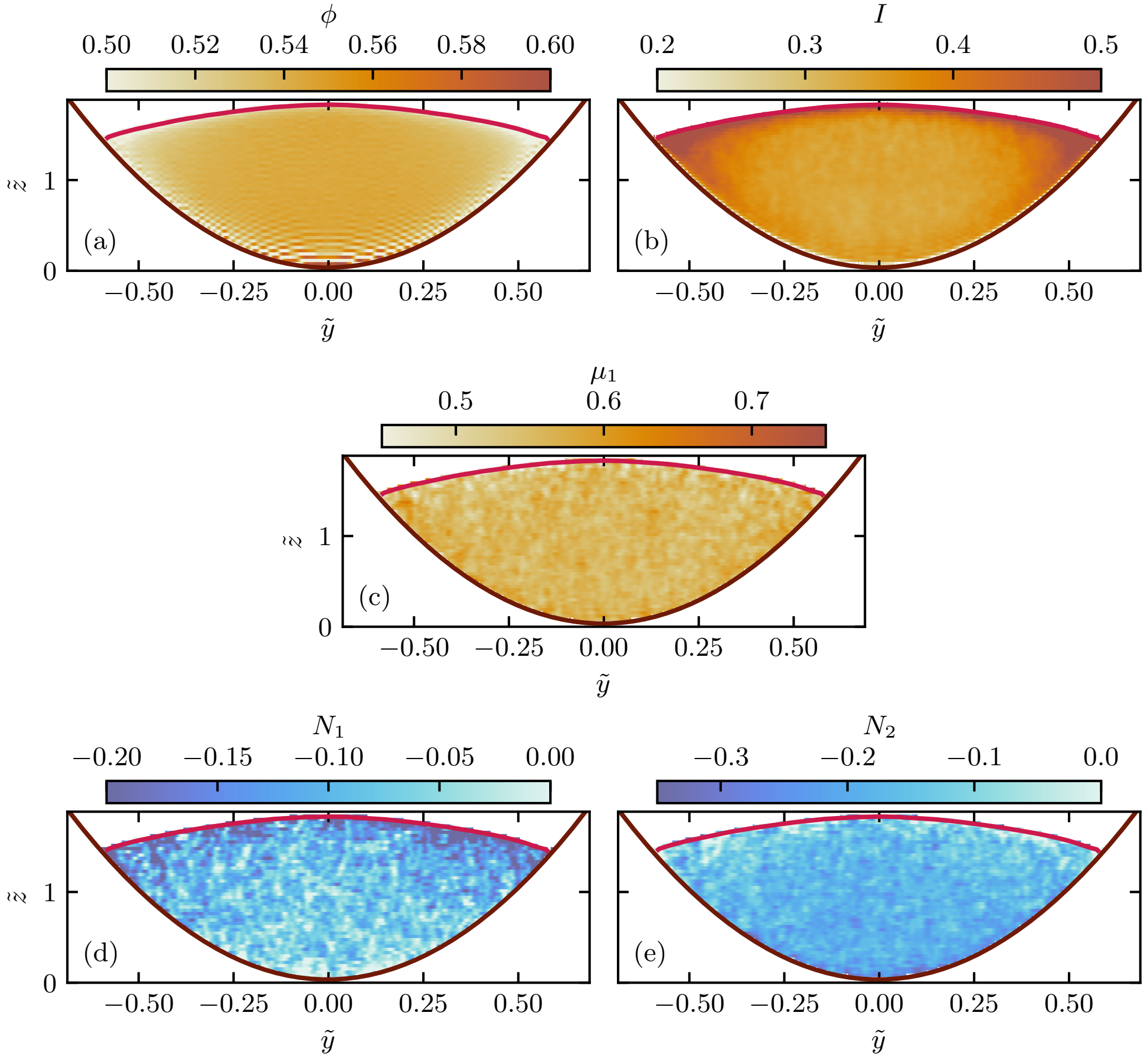

Figure 5. Uniform fields from a DEM simulation with $\theta = 29°$. (a) Particle volume fraction, $\phi$. (b) Inertial number, $I$. (c) First-order friction coefficient, $\mu_1$. (d) Scaled first normal stress difference, $N_1$. (e) Scaled second normal stress difference, $N_2$. In all subplots, the red and brown lines represent the flow surface and channel base, respectively. Note that here, the aspect ratio has been increased for visualization purposes, as the proper flow is much shallower ($\epsilon \approx 0.09$).

The volume fraction, inertial number, first-order friction coefficient, and normal stress differences can therefore be averaged spatially, and the resulting values used as an input for the continuum model to predict the surface height profile and the pressure and velocity fields. As shown by figure 6, the model and the DEM data agree quantitatively, especially for the leading order quantities, the streamwise velocity and the pressure. As shown by predictions of the leading order quantities remain accurate when changing the slope angle from 23° to 29°. However, the discrepancy between the predicted and observed velocities in the vortices increases as the slope angle decreases. This is attributed to finite-size effects ($\mathcal{H}/d$ not large enough) and further discussed in section 7.

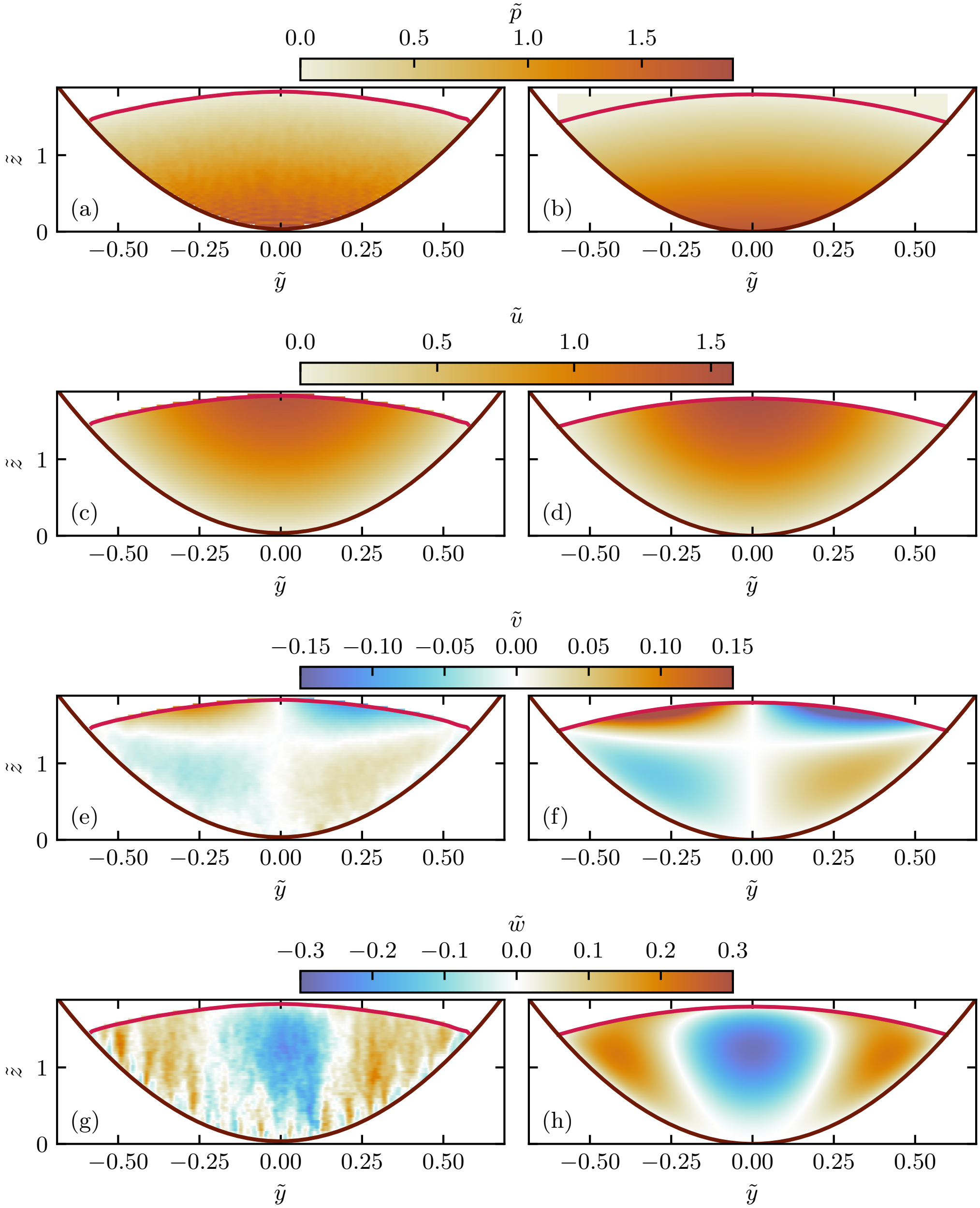

Figure 6. Comparison between dimensionless DEM results (left) and predictions from the model (right) for $\theta = 29°$ using as an input the values averaged from the uniform fields presented in figure 5. (a, b) Pressure, $\tilde{p}$, and (4.15). (c, d) Streamwise velocity $\tilde{u}$ and (4.16). (e, f) Cross-stream velocity $\tilde{v}$ and (4.25). (g, h) Normal velocity $\tilde{w}$ and (4.26). In all subplots, the red and brown lines represent the flow surface and channel base, respectively. Note that here, the aspect ratio has been increased for visualization purposes, as the proper flow is much shallower ($\epsilon \approx 0.09$).

figure 7, the prediction of the surface shape is especially accurate. Note that the cross-stream and normal velocities predicted by the model are larger than the ones observed in the DEM simulations (see discussion in section 7 for additional information). However, the model is still able to reproduce accurately the spatial structures of the vortices. The

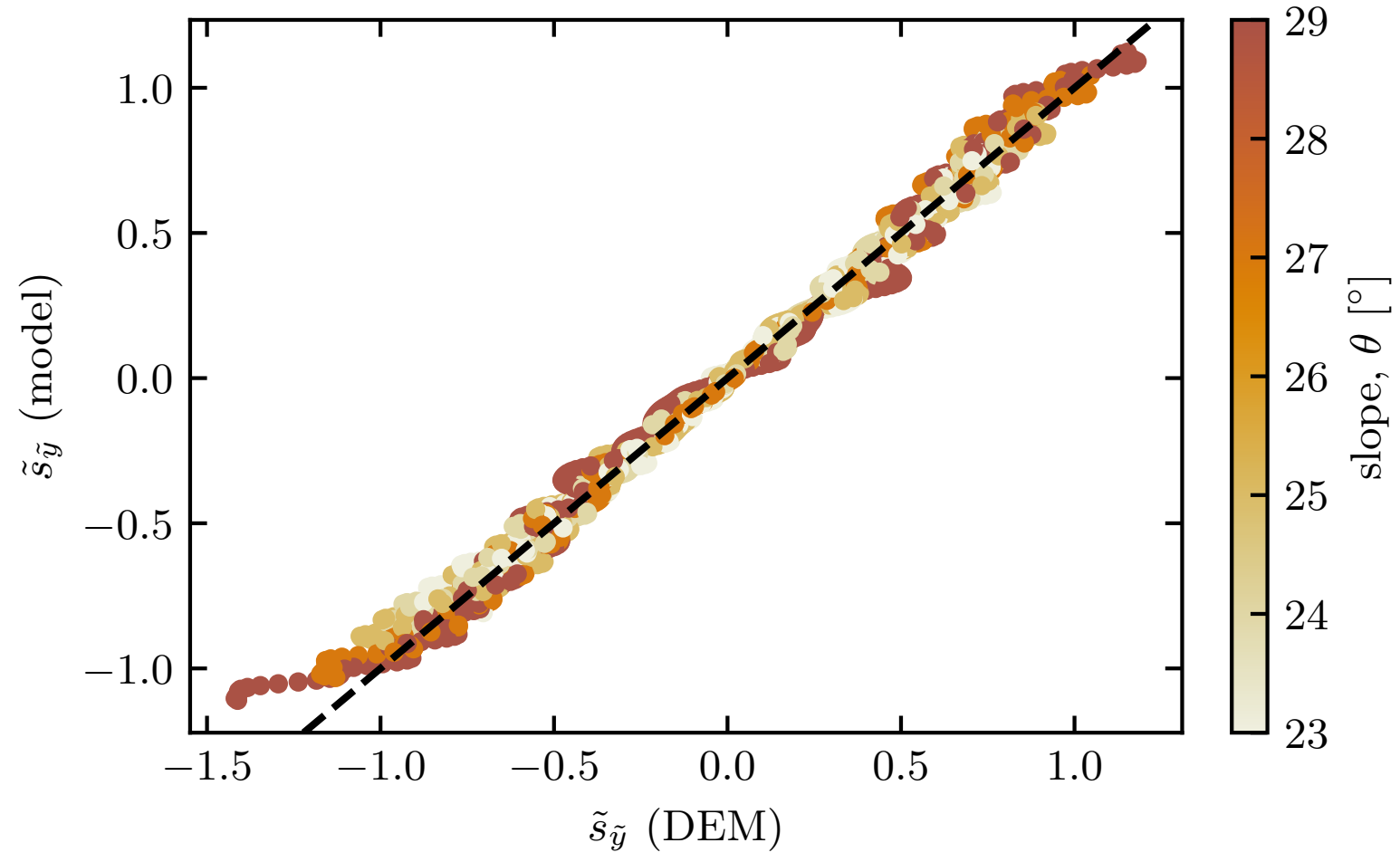

Figure 7. Comparison between dimensionless surface gradients $\tilde{s}_{\tilde{y}}$ extracted from DEM simulations and predicted from the model using as an input the values averaged from the uniform fields presented in figure 5 for three different slopes. Each point is the value of the gradient at a single position along the $y$-axis. The black dashed line is the identity line.

## 6. Laboratory experiments and inference of the second normal stress difference

An interesting result of the solution derived in section 4 is the simple linear relationship between the channel shape and the surface deformation (4.33). Hence, assuming $C$ to be unity (see the end of subsection 4.1), the second normal stress difference can be inferred from laboratory experiments by measuring the surface and basal topographies.

### 6.1. *Experimental set-up*

The experiments are carried out using a channel of curved cross-section, tilted at an angle $\theta \in [23°, 30°]$. Spherical glass beads, of diameter $d \in [125, 165]$ $\mu$m, are released from a funnel at the top of the chute behind a gate of controlled height. As they flow out of the funnel, they quickly pile up behind the door and reduce the flux coming out of the funnel, equilibrating with what comes out of the gate. This setup ensures both a constant mass flux and a controlled height at the inlet, reducing the spatial relaxation length needed for the flow to reach the gravity-friction equilibrium. The mass flux is then self-selected by the slope and the gate height, varied from 0.75 cm to 1 cm.

Height measurements are performed using the laser scanner scanCONTROL 2950-100 from Micro-Epsilon at a frequency of 120 Hz and a horizontal spatial resolution of about 0.07 mm (depending on the experiments). The vertical resolution is of the order of 10 $\mu$m, but the precision of the measurements is dominated by temporal random fluctuations, on the order of the grain diameter. The measurements are averaged during the steady part (after the passage of a front and before the running out of the flow), which lasts from 60 s to 100 s, corresponding to a total mass of about 5.7 kg of released particles. Similarly, the mass flux is recorded using a scale and averaged during the same temporal period.

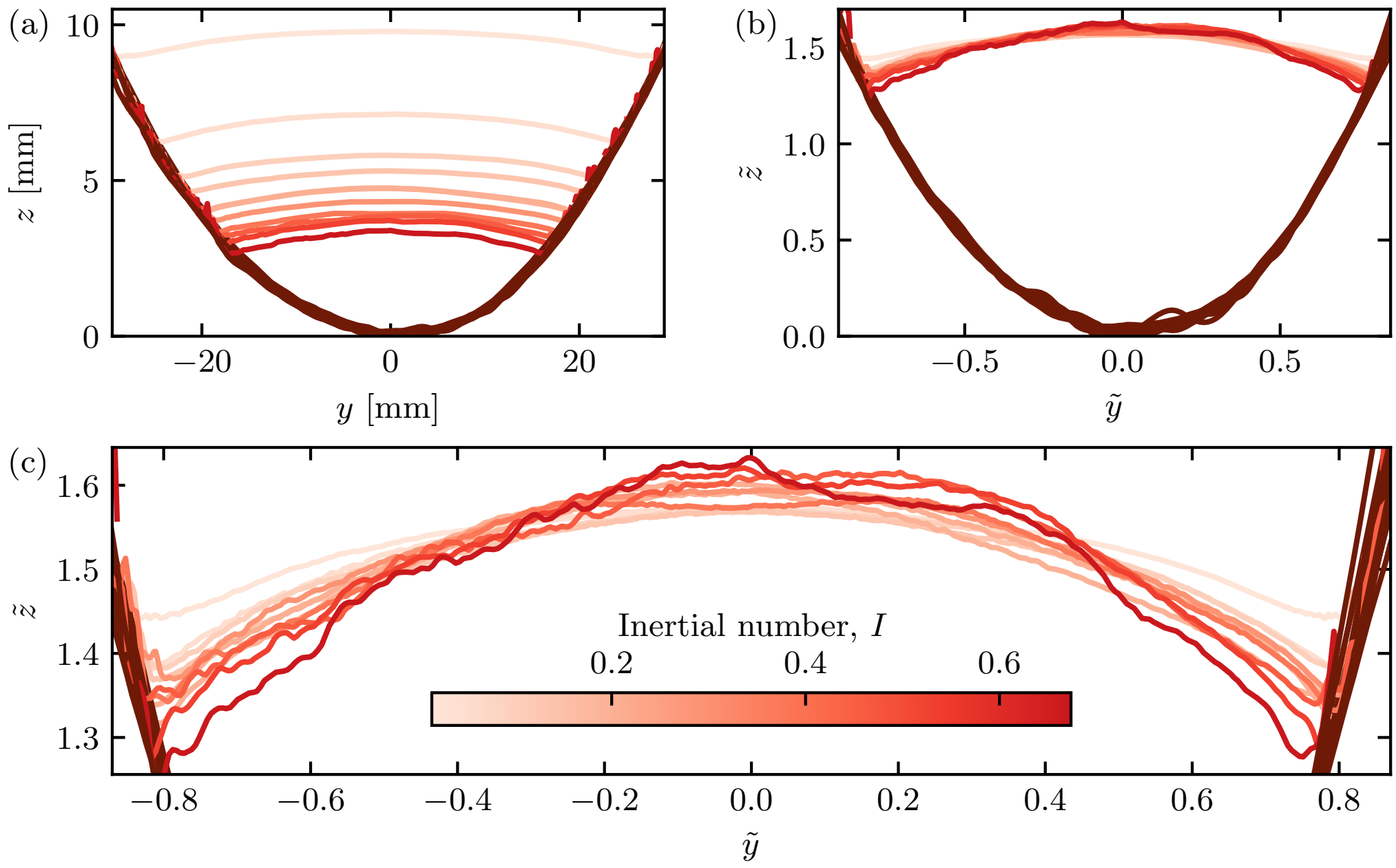

Figure 8. Surface height profiles for various slopes and fluxes in dimensional coordinates. The shades of red represent the corresponding values of the inertial number. (a) Dimensional profiles. (b) Dimensionless profiles. (c) Dimensionless profiles zoomed on the surface to highlight the evolution of the curvature with the inertial number. In all subplots, the brown curves represent measurements of the empty channel section. Not all data are shown for the sake of clarity.

### 6.2. *Surface height profiles*

As shown by figure 8a, the measured surface height profiles all exhibit an upward curvature, as observed in the DEM simulations, and predicted by (4.33) in the case of negative $N_2$. However, profiles with different flow parameters are difficult to compare, particularly across different fluxes.

The profiles can be made dimensionless using the length scales derived in sec 3.3. Their estimation, however, requires a value for the first order friction coefficient $\mu_1$, computed from the channel inclination following (4.13), and for the inertial number, which can be inferred from the mass flux measurements. Assuming the streamwise velocity follows a Bagnold profile scaled by the local flow depth (as predicted by (4.16)), using the definition of the mass flux (3.12) and integrating, the inertial number can be estimated as

$$I = \frac{5}{2} d \frac{Q}{\phi \rho_{\mathrm{p}}} \frac{1}{\sqrt{\phi g \cos(\theta) \mathcal{S}}}, \tag{6.1}$$

where

$$\mathcal{S} = \int_{-W/2}^{W/2} (s - b)^{5/2} \mathrm{d}y, \tag{6.2}$$

is computed from the experimental data.

As shown by figure 8b, all profiles collapse towards a single curve, indicating that most of the flow properties are encompassed into the gravity/friction balance used to derive the different scales. However, a careful examination of the surface shows a consistent increase of the surface curvature with the inertial number (see figure 8c).

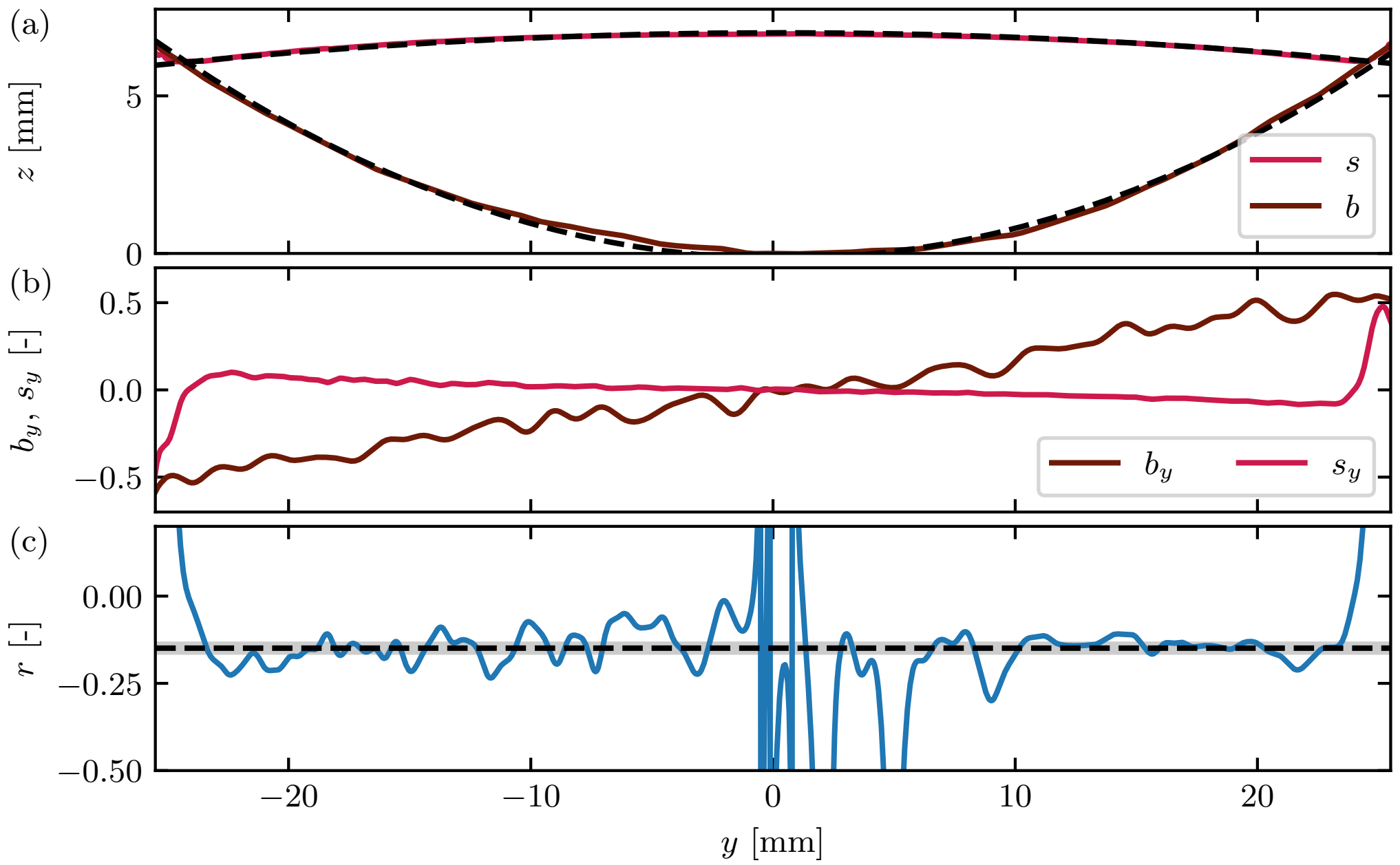

Figure 9. Experimental measurements for $\theta = 23.4°$ and $Q = 85$ g s$^{-1}$. (a) Flow surface and channel section. The black dashed lines represent the fits of (6.3) and (6.4). (b) Cross-stream gradients of the flow surface and channel section. (c) Ratio $r = s_y/b_y$. The black dashed line represents the ratio between $S/B$ from the fits in (a).

## 6.3. *Second normal stress difference*

The second normal stress difference can then be inferred from the surface height profiles. For each experimental run, the flow surface $s^{(0)}$ and channel section $b_y$ can both be approximated by a parabola (figure 9a), and their gradients $s_y^{(0)}$ and $b_y$ are found to increase linearly with $y$ (figure 9b). Furthermore, their ratio is found to be constant, as predicted by equation (4.33) (figure 9c). Hence, both are fitted with a second-order polynomial

$$s(y) = \frac{S}{2}(y - y_0)^2 + s(0), \tag{6.3}$$

$$b(y) = \frac{B}{2}(y - y_0)^2 + b(0), \tag{6.4}$$

where $S$ and $B$ are the base and flow surface gradients, and $s(0)$, $b(0)$ and $y_0$ are constants accounting for spatial shifts. Assuming $C$ to be unity (see the end of subsection 4.1), the second normal stress difference can then be computed from (4.33) as

$$N_2 = \frac{4r}{5(r-1)}, \tag{6.5}$$

where $r = s_y/b_y = S/B$.

The first order friction coefficient $\mu_1$ and the second normal stress difference $N_2$ are shown as a function of the inertial number in figure 10. The results for $\mu_1$ are in agreement with friction laws obtained experimentally in the literature with inclined plane configurations (e.g. GDR MiDi 2004). Additionally, the measurements presented here reach higher $I$-values than those performed by GDR MiDi (2004). At such values, a linear trend is found rather than a saturation, in agreement with recent measurements performed on super-stable granular heaps (Lloyd *et al.* 2025). This linear trend may result from the hypothesis of the local rheology under which all these measurements are inferred, as discussed by GDR MiDi (2004). To the authors' knowledge, no measurements of $N_2$ are available in the literature to compare with. Unfortunately, direct comparison with the results of DEM simulations is not relevant, as the results of the latter strongly depend on the particle/particle tangential friction coefficient $\mu_\mathrm{p}$ which encompasses several physical quantities, such as the particle surface roughness or geometry (Srivastava *et al.* 2020). However, $N_2$ is on the right order of magnitude, and its variation with $I$ displays similar behaviour to the ones found in numerical simulations (Nagy *et al.* 2020; Srivastava *et al.* 2020; Kim & Kamrin 2023).

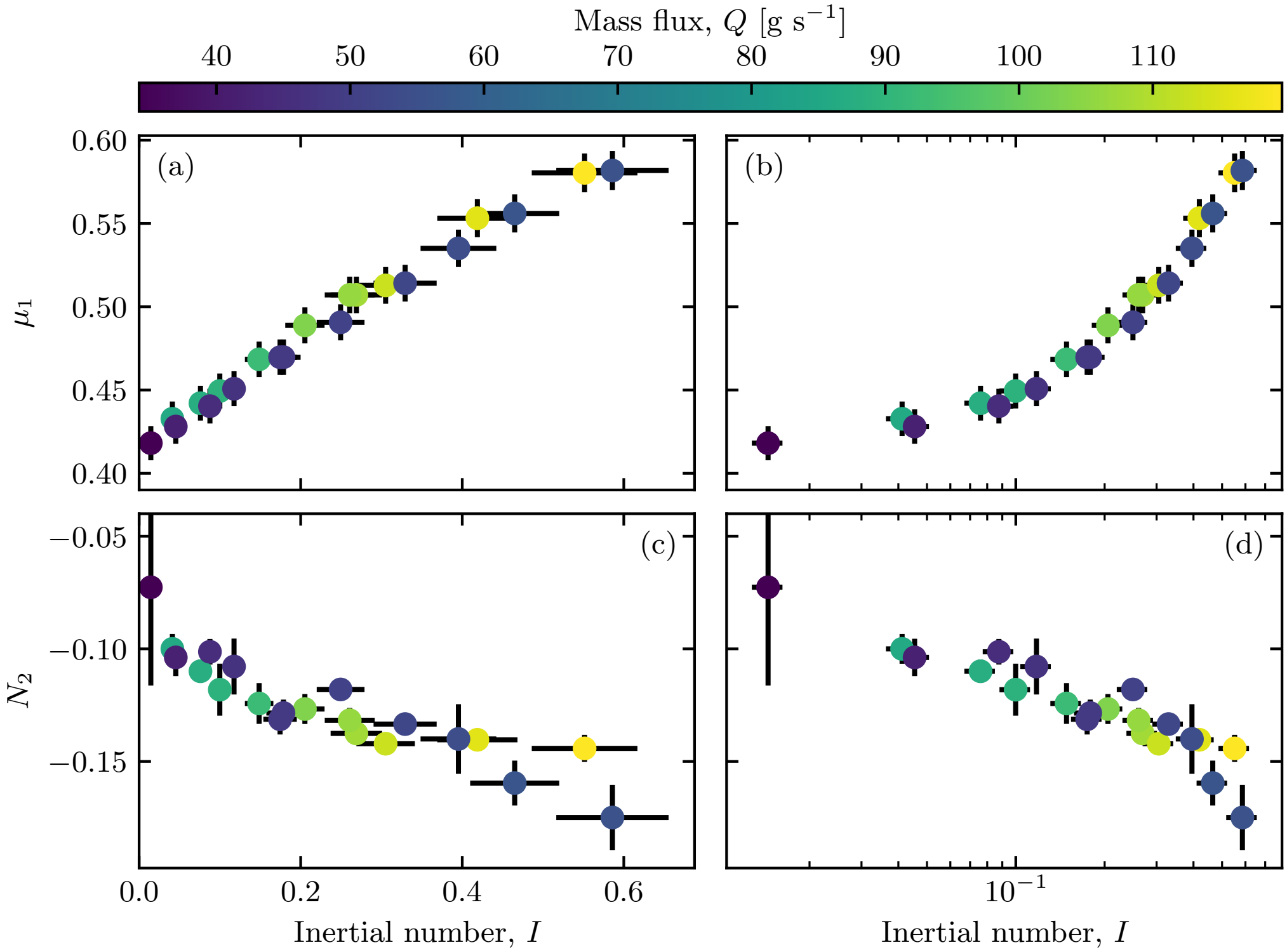

Figure 10. (a, b) Measured first-order friction coefficient as a function of the inertial number. (c, d) Measured second normal stress difference as a function of the inertial number. Note that the horizontal axes of (b) and (d) are in logarithmic scale. Errorbars represent the $\pm 1\sigma$ uncertainty.

## 7. Conclusion and perspectives

In this work, the flow of a dense granular material down an inclined channel of arbitrary cross-section is studied. Asymptotic solutions are derived from a continuum model with a second-order granular rheology accounting for the existence of normal stress differences. These solutions rely on the shallowness of the flow, which is used to expand the system of equations linearly. The resulting model can predict both the surface deformation and the existence of two counter-rotating vortices in the plane perpendicular to the flow direction, which are observed in numerical simulations and experiments in the existing literature. Importantly, the existence of these two features is linked to the presence of a cross-stream gradient of the streamwise velocity. Here, a non-flat topography is used to create this gradient, but similar results are expected in the case of a flat channel with frictional walls (McElwaine *et al.* 2012; Deboeuf *et al.* 2006; Kim & Kamrin 2023). The solution presented in this paper could be extended to this configuration, but this would require the computation of higher-order terms as the required velocity gradient would only come up at the order $\epsilon$, pushing the secondary flows to higher orders.

The validity of these solutions is tested against DEM simulations. A quantitative match is found between the two, especially concerning the prediction of the surface deformation and the structure of the vortices. The velocities involved in the secondary vortices are however smaller in the DEM than predicted by the theory. This could be attributed to the aspect ratio of the flow, not being shallow enough, but also to finite-size effects. For a fixed aspect ratio $\epsilon$, decreasing the size of the system (i.e. decreasing $\mathcal{H}/d$) also reduces the strength of the vortices up to their disappearance. This may be induced by the breakdown of the continuum assumption and/or the increase (relative to the strength of the vortices) of the self-diffusion of particles in the plane perpendicular to the flow (e.g. Utter & Behringer 2004; Artoni *et al.* 2021), but this requires a dedicated study.

An interesting feature of the predicted and observed secondary flow is the direction of the vortices, downwelling in the centre for a negative second normal stress difference. Kim & Kamrin (2023) have found vortices rotating in the opposite direction in a configuration deeper than it is wide. This suggests that the direction of the vortices could be driven not only by the sign of the normal stress difference but also by the flow aspect ratio (Lévay *et al.* 2025) or, more generally, by the shape of the container, as observed in the case of viscoelastic fluids (Yue *et al.* 2008). This interesting perspective is left for future work.

The continuum model also predicts a simple relationship between the cross-stream gradients of the surface and channel elevations. By measuring both in simple laboratory experiments, this relationship is used to infer measurements of the scaled second-normal stress difference in dense granular flows. This provides a simple method that could be used in the future to investigate the properties of the second normal-stress difference, such as its dependence on the particle properties. Importantly, similar to the $\mu(I)$ curve, the $N_2(I)$ curve is a rheological property of the granular material itself, *a priori* independent of the setup used to measure it. It can then be used as an input in continuum simulations to predict the flow of granular materials in arbitrary configurations.

**Supplementary data** Supplementary material and movies are available at [insert final doi].

**Acknowledgements** The authors acknowledge the use of open-source numerical Python libraries, including Matplotlib (Hunter 2007), NumPy (Harris *et al.* 2020), SciPy (Virtanen *et al.* 2020), which provide an incredibly efficient ecosystem allowing scientific research in Python, as well as the developers of the open-source molecular dynamics code LAMMPS used to perform the DEM simulations (Thompson *et al.* 2022). They also acknowledge the contribution of Aarnav Panda to the preliminary experiments that further lead to the experimental part of this article. They also thank Dr. Jamie Webb for useful suggestions. 

**Funding** C. Gadal acknowledges funding from the Royal Society through a Newton International Fellowship (NIF/R1/231983). J.M.N.T. Gray was supported by a Royal Society Wolfson Research Merit Award (WM150058) and an EPSRC Established Career Fellowship (EP/M022447/1). C. G. Johnson and J. M. N. T. Gray acknowledge funding from NERC through grants NE/X013936/1 and NE/X00029X/1.

**Declaration of interests** The authors report no conflict of interest.

**Data availability statement** The experimental and numerical data used in this paper are available at https://doi.org/10.5281/zenodo.15639395.

**Author ORCIDs** C. Gadal, https://orcid.org/0000-0002-2173-5837; C. G. Johnson, https://orcid.org/0000-0003-2192-3616; J.M.N.T Gray, https://orcid.org/0000-0003-3554-0499

## Appendix A. DEM simulations

The DEM simulations are run using the molecular dynamics open-source code LAMMPS (29 August 2024 stable release with update 1) (Thompson *et al.* 2022). In the model, each sphere $i$ has a diameter $d_i$ sampled within a uniform distribution ranging from $0.8d$ to $1.2d$ to prevent crystallization effects inside the flow, and a mass $m_i = \rho_s(4/3)\pi(d_i/2)^2$. Spheres interact only upon contact, which corresponds to overlapping in the model, i.e when $\delta_{ij} < 0$, where $\delta_{ij}$ is the overlapping distance:

$$\delta_{ij} = r_{ij} - \frac{d_i + d_j}{2}, \tag{A 1}$$

with $r_{ij}$ the distance between the centres of the spheres. The contact forces are modelled using a combination of a viscoelastic normal force and Coulomb tangential friction, which has proven to be efficient in modelling various granular flows (e.g. Kim & Kamrin 2023).

The viscoelastic normal force corresponds to the classical spring dashpot model (Cundall & Strack 1979):

$$F_{\rm n}^{i\to j} = \begin{cases} -\left(k_{\rm n}\delta_{ij} + \gamma_{\rm n}\dot{\delta}_{ij}\right) & \text{for} \quad \delta_{ij} < 0, \\ 0 & \text{else,} \end{cases} \tag{A 2}$$

where $k_{\rm n}$ and $\gamma_{\rm n}$ are the spring stiffness and damping coefficient, respectively. The damping coefficient is computed from the restitution coefficient $e$ as (Brilliantov *et al.* 1996):

$$\gamma_{\rm n} = \sqrt{\frac{4m_{\rm eff}k_{\rm n}}{1 + \left(\frac{\pi}{\ln e}\right)^2}}, \tag{A 3}$$

where $m_{\rm eff} = m_i m_j/(m_i + m_j)$ is the effective mass of the two colliding particles.

The tangential force corresponds to a regularized Coulomb friction also taking into account the contact history (Luding 2008):

$$F_{\rm t}^{i\to j} = -\min\left(|\gamma_{\rm t}u_{\rm t} + k_t\xi|, |\mu_{\rm p}|F_{\rm n}^{i\to j}\right), \tag{A 4}$$

where $\gamma_{\rm t}$ is the tangential damping coefficient, $u_{\rm t}$ is the tangential component of the relative velocity of the two spheres, $\xi$ the tangential displacement accumulated during the entire duration of the contact and $\mu_{\rm p}$ is the particle-particle friction coefficient. The tangential damping constant $k_{\rm t}$ is fixed to $(2/7)k_{\rm n}$ so that the frequencies of normal and tangential contact oscillations are similar, and the normal and tangential dissipation are comparable.

The channel is made of fixed particles of the same size distribution. To ensure as much as possible a no-slip boundary condition at the base of the flow, the channel is made rough by *(i)* generating a regular array of particles mapping the required channel shape, spaced of a distance $0.7d$, and *(ii)* perturbing this layer by randomly moving each sphere in every direction by a distance uniformly sampled in the range $[-0.4d, 0.4d]$. The details of this process do not matter providing the no-slip boundary condition at the base is verified.

For each particle, the particle motion is calculated by integrating Newton's second law using a Velocity-Verlet scheme with a fixed timestep $dt = 0.05t_{\rm R}$, where $t_{\rm R}$ is the Rayleigh time:

$$t_{\rm R} = \pi\sqrt{\frac{m_{\rm eff}}{k_{\rm n}}\left[1 + \ln\left(\frac{e}{\pi}\right)^2\right]}, \tag{A 5}$$

corresponding to half the period of oscillation of an undamped oscillator of the same properties. Table 1 contains the parameter values used for the simulations. The simulations are kept in the regime of *rigid* particles, where the rheology is unaffected by the particle stiffness, by ensuring that $k_{\rm n}/(Pd) > 10^4$ (Favier de Coulomb *et al.* 2017).

To save some computational time, the box is made periodic in the streamwise direction, with a length of $20d$. Depending on the exact shape of the channel, the exact flow width and maximum depth change across simulations. However, they are always around $200d$ and $30d$ (in the centre), which corresponds to about $7\times10^4$ particles per simulation.

Simulations are run for a duration of $5\times10^7 dt$, corresponding to 25 s using the values of the parameters given in table 1. Particle properties are saved every $2.4\times10^4 dt$, and are averaged using the methods described in Kim & Kamrin (2023) to obtain the continuum fields displayed in section 5.

| $d$ [mm] | $\rho_p$ [kg m$^3$] | $k_n$ [kg s$^{-2}$] | $e$ [-] | $\mu_p$ [-] |
|---|---|---|---|---|
| 0.8 | 2650 | 263000 | 0.24 | 0.4 |

Table 1. Values of the parameters used for the DEM simulations.